\documentclass[12pt]{article}
\usepackage{fullpage,amssymb,amsmath}
\usepackage[dvips]{epsfig}

\def\##1{{\bf #1}}
\def\=#1{\underline{\underline{#1}}}

\def\+#1{\underline{\bf #1}}
\def\*#1{\underline{\underline{\bf #1}}}

\def\c#1{\cite{#1}}

\def\.{\mbox{ \tiny{$^\bullet$} }}

\def\eps{\epsilon}

\begin{document}

\LARGE
\begin{center}
{\bf Negative Refraction in Outer Space?}

\vspace{10mm} \large

Tom G. Mackay\footnote{Corresponding Author. E--mail: T.Mackay@ed.ac.uk.}\\
{\em School of Mathematics,
University of Edinburgh, Edinburgh EH9 3JZ, UK}\\
  Akhlesh  Lakhtakia\footnote{Also at: Department of Physics, Imperial
College, London  SW7  2BW, UK}
\\
  {\em Department of Engineering Science and
Mechanics\\ Pennsylvania State University, University Park, PA
16802--6812, USA}

\end{center}

\vspace{4mm}

\normalsize

\section*{Negative refraction}

In the 1960s,  when the Russian physicist Victor Veselago first
speculated upon the properties of a hypothetical
isotropic material which simultaneously exhibited  negative
relative permittivity $\eps_r$  and negative relative permeability
$\mu_r$,
  the explosion of interest that was to follow almost three decades
later could scarcely have been foreseen. Amongst the many
intriguing implications  arising from $\eps_r <0$ and $\mu_r < 0
$, it is the prospect of negative refraction which has in the past
few years captured the imaginations  of  the optics and electromagnetics
research communities and inspired much lively debate \c{Rao}.

At the time Veselago's ideas concerning $\eps_r < 0$
and $\mu_r < 0$ were not widely followed up  since no
materials~---~naturally--occurring or otherwise~---~conforming to
$\eps_r < 0 $ and $\mu_r < 0$ were known to exist.
It was not until the dawn of the twenty first century
that interest in this topic was dramatically rekindled
following the first reported  experimental observation
of negative refraction  by researchers  at the
University of California, San Diego \c{Shelby}. The UCSD group tracked the
propagation of a microwave beam across the interface between air and
  a composite
metamaterial consisting of
conducting wire and ring inclusions embedded periodically on
printed circuit boards.
  Further independent experimental
reports have emerged recently which confirm the existence of microwave negative
refraction in metamaterials \c{Parazzoli,Houck}.

\section*{Negative phase velocity}

While $\eps_r < 0 $ and $\mu_r < 0$  is a sufficient condition for
negative refraction,  it is in fact not a necessary condition nor is it
strictly applicable
to real materials. In reality, materials~---~which are in essence
  collections
of charged particles~---~cannot respond instantaneously to an applied
electromagnetic field. Accordingly, all real materials are dissipative
to some degree and their constitutive parameters are complex--valued
quantities whose imaginary parts arise from dissipation.
It is readily demonstrated that the key criterion for the
negative refraction of plane waves  in real materials is that the
phase velocity be oppositely directed to the direction of energy flow
\c{McCall}.
Materials supporting such planewave propagation are called negative
phase--velocity (NPV) materials to distinguish them from conventional
positive phase--velocity (PPV) materials in which
the phase velocity and rate of energy flow are co-directional.
In the quest for negative refraction, it is significant that
NPV propagation may be predicted for materials when only one of $\eps_r$
and $\mu_r$ has a real part which is negative--valued.

With negative refraction in the microwave
regime
now appearing to be
well--established, current  efforts are  directed towards higher
frequencies \c{Yen}
with optical negative refraction being the ultimate goal.
The scope for achieving NPV propagation and thereby negative 
refraction is considerably broadened
by considering anisotropic and bianisotropic materials.
For example, through the homogenization of an isotropic chiral
material with a magnetically--biased ferrite, both of which are  PPV
materials, a
bianisotropic NPV homogenized composite may be conceptualized \c{ML_PRE}.

\section*{A relativistic perspective}

In the context of homogenous materials, we recently reported upon the exciting prospects for NPV propagation 
and negative
refraction which arise from
the Lorentz covariance of the basic laws of electromagnetics \c{NPV_STR}.
Suppose, from the perspective of an  inertial
reference
frame $\Sigma$, we have a  PPV
isotropic dielectric--magnetic material $M$.
As viewed from the perspective of another inertial frame $\Sigma'$ 
which translates
at constant velocity $v$ with respect to $\Sigma$, the material $M$ is
not an isotropic dielectric--magnetic material at all. Instead, $M$
considered from $\Sigma'$ is a non--isotropic  complex material whose
electromagnetic
constitutive
properties depend
  upon
both the orientation and magnitude of $v$. Crucially,  it has been found
that
the material $M$ can support NPV propagation provided that
  the inertial frame $\Sigma'$ from which it is observed is translating with
sufficiently high velocity. Moreover, the converse applies too: if
we have material $M$ which is an
isotropic dielectric--medium supporting NPV
propagation  in an inertial frame $\Sigma$ then it may be considered
from the perspective of an inertial frame $\Sigma'$ as a non--isotropic,
electromagnetically--complex,
PPV material provided that $\Sigma'$ translates with sufficiently high
velocity with respect to $\Sigma$.

Commonplace  terrestrial  velocities are likely to be  too low to give
rise to NPV propagation in a material which supports PPV propagation
when viewed at rest.
However, one may envisage relativistic negative
refraction being exploited  in astronomical scenarios such as, for example, in
the remote sensing of planetary and asteroidal surfaces from space stations.
Although current  research activities relating to negative refraction
are largely directed  towards the nanoscale, it may possibly be the case
that space telemetry technologies will be the first to reap the
benefits of negative refraction. Furthermore, it is possible that many
unusual phenomenons would be discovered and/or explained
by the application of the idea of relativistic negative refraction to 
interpret data
collected via telescopes. Perhaps, many more planets, hitherto
hidden, would be lit up on our space maps thereby!

\end{document}